\documentstyle[aps,twocolumn]{revtex}
\begin{document}
\twocolumn[\hsize\textwidth\columnwidth\hsize\csname@twocolumnfalse\endcsname

\title
{Feasible Linear-optics Generation of Polarization-entangled Photons Assisted with 
Single-photon Quantum Non-demolition Measurement}

\author{Zeng-Bing Chen$^1$, Jian-Wei Pan$^2$, and Yong-De Zhang$^{3,1}$}
\address
{$^1$Department of Modern Physics, University of Science and Technology of China,
Hefei, Anhui 230027, P.R. China}
\address
{$^2$Institut f\"ur Experimentalphysik, Universit\"at Wien, 
Boltzmanngasse 5, 1090 Wien, Austria}
\address
{$^3$CCAST (World Laboratory), P.O. Box 8730, Beijing 100080, P.R. China}
\date{\today}
\maketitle 

\begin{abstract}
\
We propose a feasible scheme to create $n$-party ($n\geq 2$) polarization-entangled 
photon states in a controllable way. The scheme requires only single-photon 
sources, single-photon quantum non-demolition measurement (SP-QNDM) and simple 
linear optical elements, usually with high perfections. The SP-QNDM acts as a 
non-destructive projection measurement onto the wanted entangled states and 
filters out the unwanted terms. Our scheme in fact realizes entanglement of 
non-interacting photons; the interaction occurs only implicitly in the optical 
elements and SP-QNDM. We also briefly consider purification of mixed single-photon 
states within our scheme.

\


\

\

\end{abstract}]

Einstein-Podolsky-Rosen (EPR) entanglement \cite{EPR} is one of the most
fascinating aspects of quantum mechanics (QM). Starting from local realism,
EPR concluded that QM is incomplete. Based on Bohm's \cite{Bohm}\ EPR states
Bell derived his famous inequalities \cite{Bell}, which enable to test QM
against local reality \cite{Aspect}. The conflict between QM and local
realistic theories for perfect correlations has been argued using
multiparticle entanglement [i.e., Greenberger-Horne-Zeilinger (GHZ)
correlations] \cite{GHZ-89,GHZ-90}, as has been experimentally demonstrated
recently \cite{Pan-GHZ}. In the burgeoning field of quantum information
theory \cite{QIT,Nature}, many practical applications heavily rely on
entanglement as a necessary resource. Thus the generation of entangled
states is an important task for both theoretical and practical purpose.

There are various physical systems that can be manipulated to generate
entangled states. Examples include trapped ions \cite
{trap-ion-PRL81,trap-ion-MS,trap-ion-Nature}, atom-cavities \cite
{cav-Science,cav-Guo}, Bose-Einstein condensates \cite{BEC-Nature},
optically driven Bose-Einstein condensates \cite{photon-atom} and neutral
atoms \cite{n-atom}. Among these, trapped ions and atom-cavities are two
strong candidates to implement quantum information processing (QIP) \cite
{QIT,Nature,trap-ion-CZ}{\bf . }For the purpose of testing Bell's
inequalities, the most used entanglement source is two-photon
polarization-entangled states, generated in the process of spontaneous
parametric down-conversion in a nonlinear optical crystal \cite{Aspect}.
Such a polarization entanglement plays an essential role, e.g., in quantum
teleportation \cite{QIT}. Three-photon polarization-entangled states (i.e.,
the GHZ states) have also been achieved \cite{GHZ-99}, following an earlier
proposal \cite{GHZ-PRL}. They lead to a conflict with local realistic
theories for nonstatistical predictions of QM \cite{GHZ-89,GHZ-90}, as
confirmed by a recent experiment \cite{Pan-GHZ}. Experimental four-photon
entanglement has also been reported very recently \cite{Pan-4}. GHZ states
can find practical applications in, e.g., constructing universal quantum
computation \cite{Chuang}. However, the present entangled-photon source
suffers from the low yield. This may be a severe obstacle for further
applications of entangled photons in QIP as well as in experimental study of
fundamental problems in QM.

So far, $n$-particle ($n\geq 2$) entangled states can be {\it controllably}
produced in the trapped-ion \cite{trap-ion-PRL81,trap-ion-MS,trap-ion-Nature}
and atom-cavity systems \cite{cav-Science}; the two-photon polarization
entanglement is also tunable by using a spontaneous parametric
down-conversion photon source \cite{nonmax}. Tunable entanglement may find
unprecedented applications in QIP and in the study of entanglement itself.
Nonmaximally entangled state is essential for the ``all-or-nothing'' test of
QM versus local realism in two-particle cases \cite{Hardy}. However, it is
currently unknown how to achieve the desired controllability of many-photon
entanglement.

In this paper we propose a feasible way to create $n$-party ($n\geq 2$)
polarization-entangled photon states with single-photon (SP) sources \cite
{source}, single-photon ``quantum non-demolition measurement'' (SP-QNDM) 
\cite{QND,see-1}\ and simple linear optical elements [e.g., the polarizing
beam splitter (PBS)]. The advantage of these optical elements is that their
perfection is very high, which is essential for feasible purification of
polarization entanglement \cite{Pan-2001}. Creating GHZ correlations among
particles from independent sources has been proposed previously \cite{YS}.
The recent realization of almost ideal SP turnstile device \cite{source}
enables the proposal to work in a simpler experimental arrangement and in a
broader context, as we will show below.

Now let us consider two independent SP sources (source-$1$ and source-$2$)
emitting two photons described by pure quantum states $\left| \chi
_1\right\rangle _j=\alpha _i\left| H\right\rangle _j+\beta _j\left|
V\right\rangle _j$, where $H$ ($V$) denotes the horizontal (vertical) linear
polarization and $j=1,2$; $\alpha _i$ and $\beta _i$ are complex probability
amplitudes satisfying $\left| \alpha _j\right| ^2+\left| \beta _j\right| ^2=1
$. The two photons $1$ and $2$ are incident on a PBS simultaneously. The
input mode $1$ ($2$) represents the photon $1$ ($2$). We use labels $%
1^{\prime }$ and $2^{\prime }$ to represent two output modes, in which the
output $1^{\prime }$ is subjected to a detection by a QNDM device. Since $H$
($V$) photons are transmitted (reflected) by the PBS, the output of the PBS
(with the input $\left| \chi _1\right\rangle _1\left| \chi _1\right\rangle _2
$) is 
\begin{equation}
\left| \chi _2\right\rangle =\alpha _1\alpha _2\left| H\right\rangle
_{1^{\prime }}\left| H\right\rangle _{2^{\prime }}+\beta _1\beta _2\left|
V\right\rangle _{1^{\prime }}\left| V\right\rangle _{2^{\prime }}+\left|
\chi _2^{\prime }\right\rangle   \label{two}
\end{equation}
with $\left| \chi _2^{\prime }\right\rangle =\alpha _1\beta _2\left|
0\right\rangle _{1^{\prime }}\left| HV\right\rangle _{2^{\prime }}+\alpha
_2\beta _1\left| HV\right\rangle _{1^{\prime }}\left| 0\right\rangle
_{2^{\prime }}$. Here $\left| 0\right\rangle _{j^{\prime }}$ and $\left|
HV\right\rangle _{j^{\prime }}$ denote the zero-photon and two-photon (one
with $H$-polarization and another with $V$-polarization) states along the
direction $j^{\prime }$, respectively.

Similarly to Ref. \cite{Pan-2001}, we select only ``two-mode cases'' where
there is exactly one photon in each output direction. The procedure
corresponds to a projection measurement accomplished by the QNDM device. The
form of $\left| \chi _2\right\rangle $ ensures us to infer from detections
of exactly one photon in one output mode that there is also one photon in
another output mode. In an ideal situation, this enables us to obtain, with
a probability determined by $\alpha _j$ and $\beta _j$, general two-photon
polarization-entangled states (unnormalized) 
\begin{equation}
\left| \varsigma _2\right\rangle =\alpha _1\alpha _2\left| H\right\rangle
_{1^{\prime }}\left| H\right\rangle _{2^{\prime }}+\beta _1\beta _2\left|
V\right\rangle _{1^{\prime }}\left| V\right\rangle _{2^{\prime }}
\label{detect2}
\end{equation}
if $\alpha _1\alpha _2\neq 0$ and $\beta _1\beta _2\neq 0$. When $\beta
_1\beta _2=\alpha _1\alpha _2e^{i\varphi }$ with $\varphi $ being the
relative phase between $\beta _1\beta _2$ and $\alpha _1\alpha _2$, we
generate precisely a maximally entangled state $\left| \phi _2\right\rangle
\equiv (\left| H\right\rangle _{1^{\prime }}\left| H\right\rangle
_{2^{\prime }}+e^{i\varphi }\left| V\right\rangle _{1^{\prime }}\left|
V\right\rangle _{2^{\prime }})/\sqrt{2}$ regardless an unimportant global
phase factor; otherwise $\left| \varsigma _2\right\rangle $ remain
nonmaximally entangled. Four Bell states $\left| \phi _2^{\pm }\right\rangle
_{1^{\prime }2^{\prime }}\equiv \frac 1{\sqrt{2}}(\left| H\right\rangle
_{1^{\prime }}\left| H\right\rangle _{2^{\prime }}\pm \left| V\right\rangle
_{1^{\prime }}\left| V\right\rangle _{2^{\prime }})$ and $\left| \psi
_2^{\pm }\right\rangle _{1^{\prime }2^{\prime }}\equiv \frac 1{\sqrt{2}}%
(\left| H\right\rangle _{1^{\prime }}\left| V\right\rangle _{2^{\prime }}\pm
\left| V\right\rangle _{1^{\prime }}\left| H\right\rangle _{2^{\prime }})$
can then be easily obtained by performing proper local operations. Thus by
the above procedure, one can produce entangled states in a controllable way.
Here the controllability is achieved by the SP sources, PBS and the QNDM
device, instead of spontaneous parametric down-conversion photon sources 
\cite{nonmax}.

What is crucial to our scheme is how to project $\left| \chi _2\right\rangle 
$ onto the subspace spanned by $\left| H\right\rangle _{1^{\prime }}\left|
H\right\rangle _{2^{\prime }}$ and $\left| V\right\rangle _{1^{\prime
}}\left| V\right\rangle _{2^{\prime }}$. To accomplish this, two conditions
are obviously needed. First, we require photon detectors with SP resolution.
This excludes the conventional photon detectors. Second, the measurement
must be performed in a non-destructive way; otherwise the photons after
detection will be completely destroyed and unusable. Thus the only choice is
a kind of the SP-QNDM device. Fortunately, ``seeing a single photon without
destroying it'' has already been implemented using an atom-cavity system in
a recent experiment \cite{see-1}. If by the SP-QNDM device one photon is
definitely observed along $1^{\prime }$ non-destructively, we keep the
photon, and throw it out otherwise.

To show how this works, we feed the photon(s) along $1^{\prime }$ (the
``signal'' beam) into an originally empty cavity, e.g., by a polarization
maintaining fiber. Then a ``meter'' atom (with three relevant atomic levels $%
e$, $g$, and $d$) runs into the cavity. The experiment is arranged so that
the cavity is resonant with the $e\rightarrow g$ transition and effects for
more than one photon can be ignored \cite{Grangier}. The extremely large
atom-cavity interaction (polarization-independent) entangles the meter atom
and the SP cavity mode, enabling to realize a quantum measurement. Without
going into details of the SP-QNDM experiment, it is sufficient to note the
following fact \cite{see-1,cav-Science}: After a $2\pi $ Rabi pulse, 
\begin{eqnarray}
\left| {\rm m}\right\rangle \left| \chi _2\right\rangle &\rightarrow &\left| 
{\rm m},\bar \chi _2\right\rangle =\left| {\rm m}\right\rangle _\pi (\alpha
_1\alpha _2\left| H\right\rangle _{1^{\prime }}\left| H\right\rangle
_{2^{\prime }}  \nonumber  \label{two} \\
&&+\beta _1\beta _2\left| V\right\rangle _{1^{\prime }}\left| V\right\rangle
_{2^{\prime }})+\left| {\rm m}\right\rangle \left| \chi _2^{\prime
}\right\rangle ,  \label{twoo}
\end{eqnarray}
where $\left| {\rm m}\right\rangle =c_g\left| g\right\rangle +c_d\left|
d\right\rangle $ is the initial state of the meter atom and $\left| {\rm m}%
\right\rangle _\pi =c_ge^{i\pi }\left| g\right\rangle +c_d\left|
d\right\rangle $. Thus the atom acquires a $\pi $ phase shift only if one
photon is in the cavity. As long as the $\pi $ phase shift is observed,
e.g., by atomic interferometry \cite{see-1}, $\left| \chi _2\right\rangle $
will be effectively projected onto $\left| \varsigma _2\right\rangle $
(without destroying it), according to the standard theory of quantum
measurement. After this being done, the photon is leaked out, without
changing its polarizations, as a useful resource, leaving an empty cavity
for a next SP-QNDM. It is worthwhile to point out that the required SP-QNDM
can also be accomplished by a nondestructive SP trigger \cite{trigger},
which is within the reach of current technology and plays an identical role
as the SP-QNDM based on the atom-cavity system.

The above procedure can be generalized to produce $n$-photon ($n\geq 3$)
polarization-entangled states in a ``step-by-step'' manner. This is
illustrated for the case of creating the three-photon entanglement. In this
case, we need an SP source (emitting photons in the states $\left| \chi
_1\right\rangle _1$) and two-photon entangled states. Without loss of
generality, we only consider a special input state $\left| \chi
_1\right\rangle _1\left| \psi _2^{-}\right\rangle _{23}$, where $\left| \psi
_2^{-}\right\rangle _{23}$ can be either generated by the above procedure or
from a separate EPR source. When the two photons (photon $1$ and photon $2$)
are incident on a PBS simultaneously, we make an SP-QNDM (as described
above) along any output path, selecting only the two-mode cases. Then we are
left with a three-photon entangled state 
\begin{equation}
\left| \varsigma _3\right\rangle =\alpha _1\left| H\right\rangle _{1^{\prime
}}\left| H\right\rangle _{2^{\prime }}\left| V\right\rangle _3-\beta
_1\left| V\right\rangle _{1^{\prime }}\left| V\right\rangle _{2^{\prime
}}\left| H\right\rangle _3,  \label{detect3}
\end{equation}
which has been normalized. Tuning $\alpha _1$ and $\beta _1$ in $\left| \chi
_1\right\rangle _1$ enables us to control the entanglement of the produced
state $\left| \varsigma _3\right\rangle $. Particularly, when $\left| \alpha
_1\right| =\left| \beta _1\right| $, $\left| \varsigma _3\right\rangle $ is
just a maximally entangled three-photon state, i.e., a GHZ state observed
experimentally in Ref. \cite{GHZ-99}. In principle, we can also accomplish
the same task in a single step by using three independent SP sources and two
PBS put in succession. Generally speaking, to generate $n$-photon ($n\geq 2$%
) entangled states, one requires at most $n$ SP sources, $n-1$ PBS and the
SP-QNDM performed $n-1$ times (other optical elements such as wave plates
are also needed, of course, so that we can perform appropriate local unitary
operations on the polarization qubits when needed).

Note that in the above scheme, the coincidence detection was not exploited
since it is equivalent to ensure that each source emits only one photon \cite
{Pan-4}, which is almost perfectly realized with current SP source \cite
{source}. Even when the coincidence detection is needed, the two-fold one is
sufficient since only the two-mode cases are involved here.

Now it is clear that in our scheme, the SP-QNDM acts as a nondestructive
projection measurement onto the wanted entangled states and filters out the
unwanted terms. Taking Eq. (\ref{detect3}) as an example, the SP-QNDM will
select $\left| H\right\rangle _{1^{\prime }}\left| H\right\rangle
_{2^{\prime }}\left| V\right\rangle _3$ and $\left| V\right\rangle
_{1^{\prime }}\left| V\right\rangle _{2^{\prime }}\left| H\right\rangle _3$
simultaneously, while preserving their coherence. As comparison, in the
proposal of generating the GHZ states \cite{GHZ-99,GHZ-PRL}, one uses the
four-fold coincidence detection, which can select $\left| H\right\rangle
_{1^{\prime }}\left| H\right\rangle _{2^{\prime }}\left| V\right\rangle _3$
and $\left| V\right\rangle _{1^{\prime }}\left| V\right\rangle _{2^{\prime
}}\left| H\right\rangle _3$ only separately and at the same time, performs a
specific (destructive) measurement on the resulting entangled states. To
retain the coherence, the quantum erasure technique is necessary.

A disadvantage of the scheme proposed here is that the entangled states are
produced in a non-deterministic way, i.e., one succeeds only with
probability less than one. This drawback is, however, underlying most
proposals of generating entangled states, with an exception in the
trapped-ion system \cite{trap-ion-PRL81,trap-ion-MS,trap-ion-Nature}. Note
that in an earlier proposal \cite{GHZ-PRL}, producing a GHZ state needs two
pairs of maximally entangled photons and the fourfold coincidence detection.
Thus the overall efficiency of the proposal is rather low due to the low
generating rate of current entangled-photon sources and low efficiency of
the four-fold coincidence detection \cite{GHZ-99}. By contrast, the present
scheme uses the SP sources and the SP-QNDM in each step. Thus a remarkable
advantage of our scheme is that {\it we do not need the pre-existing
entanglement resource; in some sense, the PBS and SP-QNDM create the
``expensive'' resource on demand}. When the SP sources are bright enough,
our scheme, despite its non-deterministic nature, may enhance the efficiency
substantially by performing the SP-QNDM as fast as possible, resulting in a
high generating rate of entangled photons. To produce multiparticle
entangled states, one can, in principle, use also the quantum controlled-NOT
(or similar) operations \cite{QIT}. Unfortunately, these quantum logic
operations are very difficult to realize with current experimental
technology. A recent entanglement purification protocol \cite{Pan-2001}
replaces the quantum controlled-NOT with PBS and is thus feasible. Compared
with Ref. \cite{Pan-2001}, the feasibility of our scheme is apparent since
the SP-QNDM is realizable already.

In the above proof-of-principle argument, we consider only ideal situations
(e.g., pure SP states and ideal SP-QNDM). Practically, there are several
detrimental factors that limit the efficiency of our scheme and must be
carefully taken into account. In the following we merely discuss the cases
when the SP states are mixed, leaving other limiting factors for future
work. In a realistic experiment, SP states will, in general, carry
correlation information of other degrees of freedom (e.g., time). These
correlations must be erased since any of them might in principle be used to
destroy the coherence of photons. In particular, the photons incident on the
PBS must be time synchronized correctly (e.g., within a time window of the
coherent time of the SP sources) so that any which-way information is erased 
\cite{QIT,swap}. Here we make a further observation that in essence, these
correlations necessarily drive the pure SP states $\left| \chi
_1\right\rangle $ into mixed ones. Thus the problem reduces to how to purify
generally mixed SP states.

Fortunately, purification of mixed SP states is easy to implement by
exploiting the entanglement purification protocol developed recently \cite
{Pan-2001}. But now only the SP sources, PBS and QNDM (instead of the
quantum controlled-NOT operations \cite{QIT}) are needed as above. For
simplicity, consider an ensemble of photons described by the density matrix
of special form (the restriction to this special case is not essential) 
\begin{equation}
\rho =f\left| H\right\rangle \left\langle H\right| +(1-f)\left|
V\right\rangle \left\langle V\right| .\;\;\;(1>f>\frac 12)  \label{mix}
\end{equation}
As usual, we divide the ensemble into two equal subensembles $\rho _1$ and $%
\rho _2$, which are described by the same density matrix (\ref{mix}). Then
two photons, one taken from the subensemble $1$ and another from
subensembles $2$ are superimposed on a PBS simultaneously. After the PBS one
selects again the two-mode cases to obtain the ensemble 
\begin{eqnarray}
\rho _{12} &=&f^2\left| H\right\rangle _{1^{\prime }}\left\langle H\right|
\otimes \left| H\right\rangle _{2^{\prime }}\left\langle H\right|  \nonumber
\\
&&+(1-f)^2\left| V\right\rangle _{1^{\prime }}\left\langle V\right| \otimes
\left| V\right\rangle _{2^{\prime }}\left\langle V\right| .  \label{mix2}
\end{eqnarray}
The fact that $\rho _{12}$ is entangled implies that the procedure also
prepares mixed entangled states. Next, we introduce a new basis spanned by $%
\left| \pm \right\rangle =\frac 1{\sqrt{2}}(\left| H\right\rangle \pm \left|
V\right\rangle )$, which enables us to rewrite 
\begin{eqnarray}
\rho _{12} &=&[f^2\left| H\right\rangle _{1^{\prime }}\left\langle H\right|
+(1-f)^2\left| V\right\rangle _{1^{\prime }}\left\langle V\right| ] 
\nonumber \\
&&\otimes \frac 1{\sqrt{2}}(\left| +\right\rangle _{2^{\prime }}\left\langle
+\right| +\left| -\right\rangle _{2^{\prime }}\left\langle -\right| ) 
\nonumber \\
&&+[f^2\left| H\right\rangle _{1^{\prime }}\left\langle H\right|
-(1-f)^2\left| V\right\rangle _{1^{\prime }}\left\langle V\right| ] 
\nonumber \\
&&\otimes \frac 1{\sqrt{2}}(\left| +\right\rangle _{2^{\prime }}\left\langle
-\right| +\left| +\right\rangle _{2^{\prime }}\left\langle -\right| ).
\label{newb}
\end{eqnarray}
Then perform polarization measurement at $2^{\prime }$ in the $+/-$ basis.
According to the measurement results, one can always obtain, by proper local
operations and classical communication, a purified ensemble 
\begin{equation}
\rho ^{\prime }=f^{\prime }\left| H\right\rangle _{1^{\prime }}\left\langle
H\right| +(1-f^{\prime })\left| V\right\rangle _{1^{\prime }}\left\langle
V\right|  \label{mixp}
\end{equation}
with a larger fraction $f^{\prime }=f^2/[f^2+(1-f)^2]>f$ of photons in the
state $\left| H\right\rangle _{1^{\prime }}$. As long as the initial
ensemble is sufficiently large, iterating the procedure will yield a state
of photons arbitrarily close to $\left| H\right\rangle _{1^{\prime }}$, from
which we are able to get any desired pure state by local unitary operations.
This concludes our discussion on purifying mixed states of the form (\ref
{mix}). The extension of the discussion to more general cases is also
possible.

Note that though the above purification protocol is developed here for a
specific purpose, it is interesting in its own right. It may be important
for applications in QIP. For instance, the preparation of arbitrary initial
state as an input is important for quantum computing. The purification
protocol is also a feasible way to suppress the decoherence of qubits,
necessarily caused by uncontrollable interaction between the qubits and
their surroundings. More importantly, purifying entanglement is also
feasible within our scheme.

Usually, entangling two particles requires interaction between them, which
is difficult to implement for photons. The above discussion in fact realizes
entanglement of non-interacting photons; the interaction occurs only
implicitly in the PBS and QNDM. Remarkably, efficient quantum computation
with linear optics is also possible, realizing the dream of computing with
non-interacting particles \cite{Knill}. Put together, it leads to the
expectation that linear optical elements might play a fundamental role in
future development of quantum information applications. The main obstacle to
scalable optical QIP lies in the requirement for nonlinear couplings between
optical modes of few photons, a difficult experimental task \cite{Knill}.
The obstacle seems to be absent here: The ability to manipulate entanglement
with the PBS assisted with the SP-QNDM might imply the usefulness of our
scheme in implementing quantum computing, though in a non-deterministic way.

To summarize, we have suggested a feasible (and perhaps even efficient)
scheme to prepare polarization-entangled photon states, which act as a kind
of an expensive resource for quantum computational and communicational
tasks. As is widely believed, the photonic channel is essential for quantum
informational applications since photons are well suited for transmitting
quantum information over long distances \cite{QIT,photon-atom,Zoller}. Thus
the present scheme, together with the ability to purify mixed SP and
polarization-entangled states, might be a crucial ingredient for both
theoretical and practical applications. Combined with recent advances in
cavity quantum electrodynamics \cite{see-1,cav-Science,trap-1}, one might
envisage a quantum network based on atom-cavities (performing SP-QNDM) and
PBS (creating polarization entanglement of photons): Atoms in the cavity as
quantum memory and photons as transmitters of quantum information \cite
{QIT,Zoller}. In this picture, the proposed scheme is not only necessary but
also natural.

This work was supported by the National Natural Science Foundation of China
under Grant No. 19975043.

\end{document}